\documentclass[aps,
               pra,
               twocolumn,
               nofootinbib,
               floatfix,
               showpacs
              ]{revtex4}

\usepackage[pdftex]{graphicx}         % fuer das Einbinden von Grafiken
\usepackage{amsmath}
\usepackage{amscd}
\usepackage{amsfonts}
\usepackage{amsthm}
\usepackage{amssymb}
\usepackage[bookmarksopen]{hyperref}
\usepackage{natbib} 
\usepackage{color}
\usepackage{setspace}
\usepackage{booktabs}
\usepackage{multirow}
\usepackage{slashbox}
\usepackage{bm}
\usepackage{units}
\usepackage{upgreek}
\definecolor{darkblue}{rgb}{0.0,0.0,0.4}
\definecolor{darkgreen}{rgb}{0.0,0.4,0.0}
\definecolor{darkred}{rgb}{0.6,0.0,0.0}

\renewcommand{\i}{\mathrm{i}}

\newcommand{\eq}[1]{(\ref{eq:#1})}
\newcommand{\Eq}[1]{Eq.~(\ref{eq:#1})}

\newcommand{\Fig}[1]{Fig.~\ref{fig:#1}}

\newcommand{\Sectionref}[1]{Section~\ref{sec:#1}}
\newcommand{\Sect}[1]{Sect.~\ref{sec:#1}}

\hypersetup{	 colorlinks,linkcolor=darkgreen,citecolor=darkred,urlcolor=darkblue, 
		pdftitle = {Momentum spectra of soliton ensembles in a one-dimensional Bose gas},
		pdfauthor = {Thomas Gasenzer},
		pdfsubject = {Solitons, 1D, bose gases, nonequilibrium}	
	    }

\begin{document}

% declarations for front matter
\title{Nonthermal fixed points and solitons in a one-dimensional Bose gas}
\author{Maximilian Schmidt}
\author{Sebastian Erne}
\author{Boris Nowak}
\author{D\'enes Sexty}
\author{Thomas~Gasenzer}
\email{t.gasenzer@uni-heidelberg.de}
\affiliation{Institut f\"ur Theoretische Physik,
             Ruprecht-Karls-Universit\"at Heidelberg,
             Philosophenweg~16,
             69120~Heidelberg, Germany}
\affiliation{ExtreMe Matter Institute EMMI,
             GSI Helmholtzzentrum f\"ur Schwerionenforschung GmbH, 
             Planckstra\ss e~1, 
             64291~Darmstadt, Germany} 

\date{\today}

\begin{abstract}
Single-particle momentum spectra for a dynamically evolving one-dimensional Bose gas are analysed in the semi-classical wave limit.
Representing one of the simplest correlation functions these give information about possible universal scaling behaviour.
Motivated by the previously discovered connection between (quasi-)topological field configurations, strong wave turbulence, and nonthermal fixed points of quantum field dynamics, soliton formation is studied with respect to the appearance of transient power-law spectra.
A random-soliton model is developed to describe the spectra analytically, and the analogies and difference between the appearing power laws and those found in a field theory approach to strong wave turbulence are discussed.
The results open a view on solitary wave dynamics from the point of view of critical phenomena far from thermal equilibrium and on a possibility to study this dynamics in experiment without the necessity of detecting solitons in situ.
\end{abstract}


\maketitle

%==================================================================
\section{Introduction}
\label{sec:intro}
Many-body systems far away from thermal equilibrium can show a much wider range of characteristics than equilibrium systems.
However, they buy this abundance for the price of transiency.
Among the wealth of possible non-equilibrium many-body configurations most interesting candidates for theoretical and experimental study are those at which generic time-evolutions get stuck for an extraordinarily long time.
This is in particular possible at critical points where universal properties like power-law behaviour of correlation functions appear, leading to slowing-down phenomena as infrared modes become dominant.
Out of equilibrium, fluid turbulence is among the earliest described as a critical phenomenon \cite{Kolmogorov1941a, Obukhov1941a,Frisch1995a, Eyink1994a}.

In the realm of quantum physics,  far-from-equilibrium many-body dynamics, from the formation of Bose-Einstein condensates in ultracold gases to quark-gluon plasmas produced in heavy-ion collisions and reheating after early-universe inflation, exhibits many interesting phenomena.
In this context, increasing attention \cite{Micha:2002ey,Berges:2008wm,Berges:2008sr, Berges:2008mr,Scheppach:2009wu,Berges:2010ez,Arnold:2005ef,
%Arnold:2005qs,
Mueller:2006up,Carrington:2010sz,Fukushima2011a,Nowak:2010tm,Gasenzer:2011by,Nowak:2011sk,Berges:2012us} is being given to wave turbulence phenomena \cite{Zakharov1992a,Nazarenko2011a,Newell2011a}.
Critical points far from equilibrium, so-called nonthermal fixed points, were proposed \cite{Berges:2008wm} and related to strong wave turbulence \cite{Berges:2008sr, Berges:2008mr,Scheppach:2009wu,Berges:2010ez} and the formation of quasi-topological defects \cite{Nowak:2010tm,Gasenzer:2011by,Nowak:2011sk}.
Such defects play an important role in superfluid turbulence, also referred to as quantum turbulence (QT), which has been the subject of extensive studies in the context of helium \cite{Halperin2008a, Donnelly1991a} and dilute Bose gases~\cite{Horng2007a,Horng2009a,Foster2010a,Yukalov2010a}. 
In contrast to eddies in classical fluids, vorticity in a superfluid is quantised \cite{Onsager1949a,Feynman1955a}, and the creation and annihilation processes of quantised vortices are distinctly different \cite{Halperin2008a,Donnelly1991a}. 

Considering a single vortex or vortex line, geometry imposes a power-law shape of the angle-averaged flow velocity away from the vortex core.
This spatial power-law dependence implies the power-law momentum spectrum predicted to occur at the nonthermal fixed point \cite{Nowak:2010tm,Nowak:2011sk}.
In this article we focus on solitary waves in one-dimensional quantum gases and show that these exhibit scaling behaviour in analogy to the universal properties of vortex ensembles in two- and three-dimensional systems.
Self similarity here is due to the absence of a scale in a soliton which has zero width and is randomly positioned in space.
See Refs.~\cite{Rajantie:2006gy,
%Rajantie:2010tb,
Berges:2010nk} for related studies in relativistic field theory.

Solitons are non-dispersive wave solutions which can arise in many non-linear systems spanning a wide range from the earth's atmosphere \cite{Roger} or water surface waves \cite{Scheffers} to optics \cite{Mollenauer}. 
The characteristics of single or few solitons in ultracold Bose gases have been studied during the last decade regarding their movement and interaction in traps \cite{Theocharis2007,Oberthaler2008, Becker2008, Scott2010}, their formation and creation \cite{Zurek2010,Zurek2010-1,Burger2002,Carr2001} and their decay \cite{Fedichev1999,Busch2000}, see Ref.~\cite{Frantzeskakis} for a review.
Here, we propose to detect and characterise soliton excitations in ultracold gases by measuring the single-particle momentum spectra for such systems.
We emphasise that these can give a strong indication for the presence of solitons even in cases where they cannot be observed in situ.

Superfluid turbulence plays an important role in the context of the kinetics of condensation and the development of long-range order in a dilute Bose gas. 
This, as well as turbulence in its acoustic excitations has been discussed in Refs.~\cite{Levich1978a,Kagan1992a,Kagan1994a,
%Kagan1997c,
Berloff2002a,Svistunov2001a}, and more recently in Refs.~\cite{Kozik2009a,Nowak:2010tm,Nowak:2011sk,Berges:2012us}.
A possible observation of QT in ultracold atomic gases presently poses an exciting task for experiments \cite{Weiler2008a,Henn2009a,seman2011}. 
We stress that the experimental study of superfluid turbulence and, more generally, of nonthermal fixed points in ultracold Bose gases is central for the understanding of the processes important during the build-up of coherence and degeneracy in particular also in one-dimensional systems~\cite{hofferberth2007, Kitagawa2011,Armijo2011a}.

In the following we study the formation of soliton excitations in trapped one-dimensional Bose gases by means of simulations in the classical-wave limit of the underlying quantum field theory. 
We analyse the momentum spectra of the time-evolving system and discuss them with respect to nonthermal fixed points discussed in field theory. 
In \Sectionref{SPMS}, we develop a model of independent, randomly positioned grey solitons, locally being solutions of the Gross-Pitaevskii equation describing a degenerate, one-dimensional dilute Bose gas.
Single-particle momentum spectra are derived for such systems, both in a homogeneous system and within a trapping potential.
A protocol for the formation of such soliton ensembles within a trapped gas is described in \Sectionref{Simulations}, and the resulting states are analysed with respect to the predicted momentum spectra and power-law signatures of a nonthermal fixed points.
Our conclusions are drawn in \Sectionref{Conclusions}.

%
%------------------------------------------------------------------------------------------
 \begin{figure}[!t]
 \includegraphics[width=0.48\textwidth]{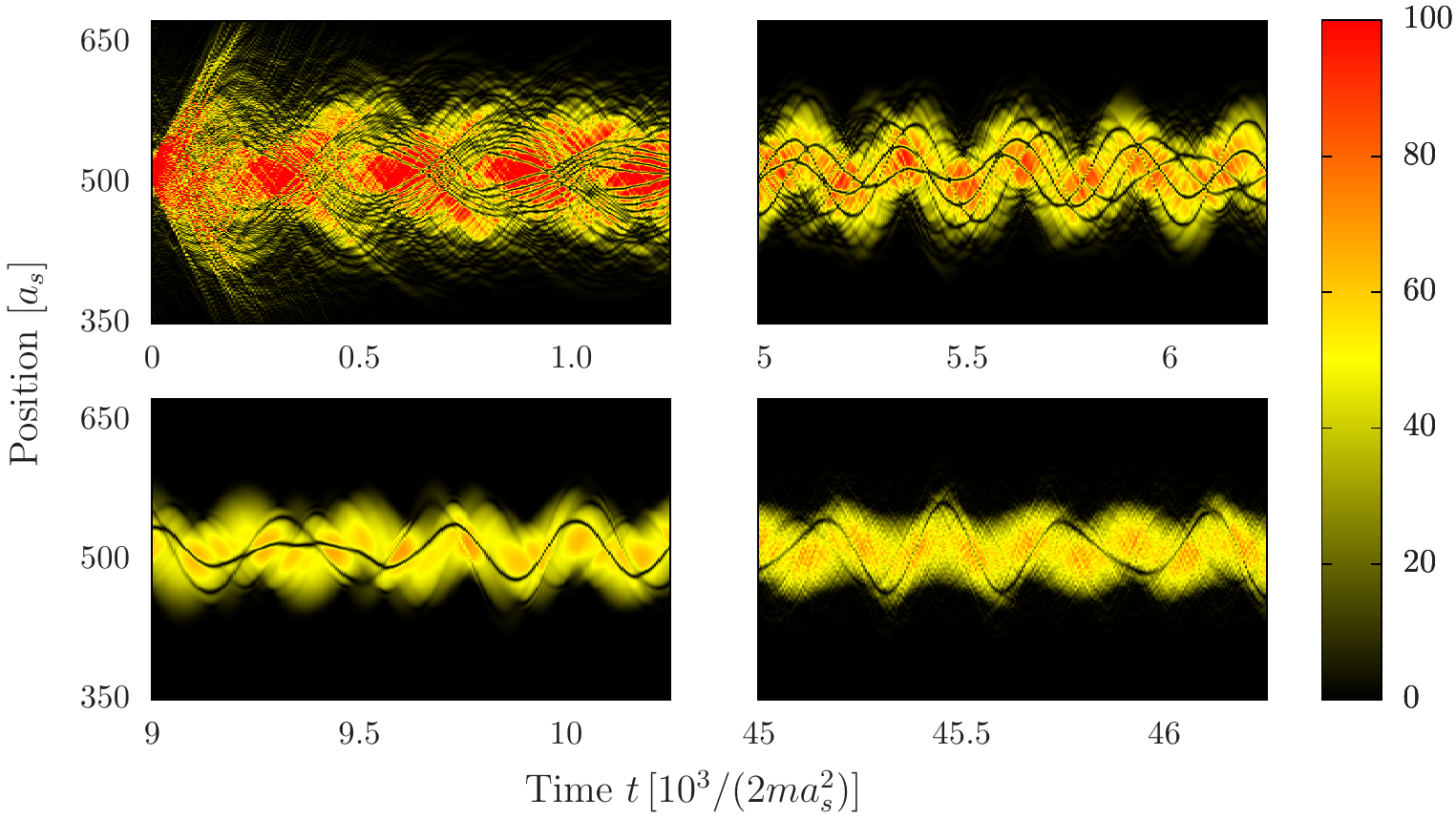}
 \caption{Snapshots of a single run of the non-linear classical field equation, showing solitons which oscillate inside a trapped one-dimensional ultracold Bose gas. 
The gas is initially noninteracting and thermalised, with $T=360\omega_{\mathrm{ho}}$, in a trap with oscillator length $l_{\mathrm{ho}}=8.5$ (in grid units). 
At time $t=0$ the interaction is switched to $g_{\mathrm{1D}}=7.3\times10^{-3}$, and a cooling period using a high-energy knife is applied, see \Sect{Simulations} for details on grid units, the chosen parameters and protocol.
The panels show the one-dimensional colour-encoded density distribution as a function of time. 
Top left panel: The initially imposed interaction quench causes strong breathing-like oscillations and the creation of many solitons. 
Top right panel: Breathing oscillations have damped out, leaving a dipolar oscillation of the bulk distribution in the trap. 
Clearly distinct solitons have formed. 
Bottom left panel: After the end of the cooling period (here at $t=t_{c}=9.1\times10^{3}$) only a few solitons are left oscillating within the oscillating bulk.
Bottom right panel: A single soliton is left at late times.}
 \label{fig:RSPACEevolution}
 \end{figure}
%------------------------------------------------------------------------------------------

%==================================================================
\section{Momentum spectra of soliton ensembles}
\label{sec:SPMS}
In the classical-wave limit, a dilute ultracold Bose gas is well described by a positive definite Wigner phase-space distribution function $W[\phi,\phi^{*}]$ for the complex field $\phi$ at each point in position or momentum space \cite{polkovnikov2010, blakie2008}.
The dynamics of the gas is determined by the time evolution of this Wigner function according to the classical field equation 
\begin{equation}
\label{eq:GPE}
 \i \partial_t \phi(x,t) = \left(-\frac{{\partial_x}^2}{2m}  + V(x,t) + g_\mathrm{1D} \left| \phi(x,t) \right| ^2 \right)  \phi(x,t)
\end{equation}
which is identical in form to the Gross-Pitaevskii equation (GPE) for the field expectation value.
Here, $m$ is the mass of the bosons, $V$ an external trapping potential, and $g_{\mathrm{1D}}$ the coupling constant in one spatial dimension (1D).

The 1D Gross-Pitaevskii equation \eq{GPE} possesses quasi-topological soliton solutions which may travel with a fixed velocity but are non-dispersive, i.e., stationary in shape \cite{Frantzeskakis}. 
For positive coupling constant $g_\mathrm{1D}>0$ solitons are characterised by an exponentially localised density depression with respect to the surrounding bulk matter and a corresponding shift in the phase angle $\varphi$ of the complex field $\phi=|\phi|\exp\{i\varphi\}$. 
Depending on the depth of this depression, the soliton is either called grey or, for maximum depression, black.
On the background of a homogeneous bulk density $n$ it is described by
\begin{equation}
\label{eq:SingleSoliton}
\phi_{\nu}(x,t)=\sqrt{n}\left[\gamma^{-1}\tanh\left(\frac{x-x_{s}(t)}{\sqrt{2}\gamma\xi}\right)+i\nu\right]
\end{equation}
where $x_{s}(t)=x_{0}+\nu t$ is the position of the soliton at time $t$.
Here, $\xi=[2mng_{\mathrm{1D}}]^{-1/2}$ is the healing length, and we express time in units of the inverse speed of sound $c_{s}=[ng_{\mathrm{1D}}/m]^{1/2}$, i.e., $t=\bar t/c_{s}$, and drop the overbars.  
$\gamma=1/\sqrt{1-\nu^{2}}$ is the `Lorentz factor' corresponding to the velocity $v$ of the grey soliton in units of the speed of sound, $\nu=v/c_{s}=|\phi_{\nu}(vt,t)|/\sqrt{n}$.
Being related to the density minimum, $\nu$ is also termed the `greyness' of the soliton,  ranging between $0$ (black soliton, $|\phi_{\nu}(vt,t)|=\nu\sqrt{n}=0$) and $1$ (no soliton, $|\phi_{\nu}(vt,t)|=\sqrt{n}$).

In \Sect{Simulations} we will study the formation and evolution of soliton excitations in trapped one-dimensional Bose gases by means of the GPE \eq{GPE}.
An example of one run is shown in \Fig{RSPACEevolution}: 
A sudden initial quench of the coupling $g_\mathrm{1D}$ creates strong oscillations of the bulk gas in the trap and lets solitons form out of the short-wave-length collective oscillations.
These solitons are seen as black lines surviving within the oscillating gas for a long time. 

In this article we are mainly interested in the characterisation of the ensemble of solitons emerging in our simulations in terms of single-particle spectra in momentum space.
Hence we refer to \Sect{Simulations} for more details on the simulation protocol and results and start here with a discussion of the possible spectra.

\begin{figure}[t]
 \centering \includegraphics[width=1.0 \linewidth]{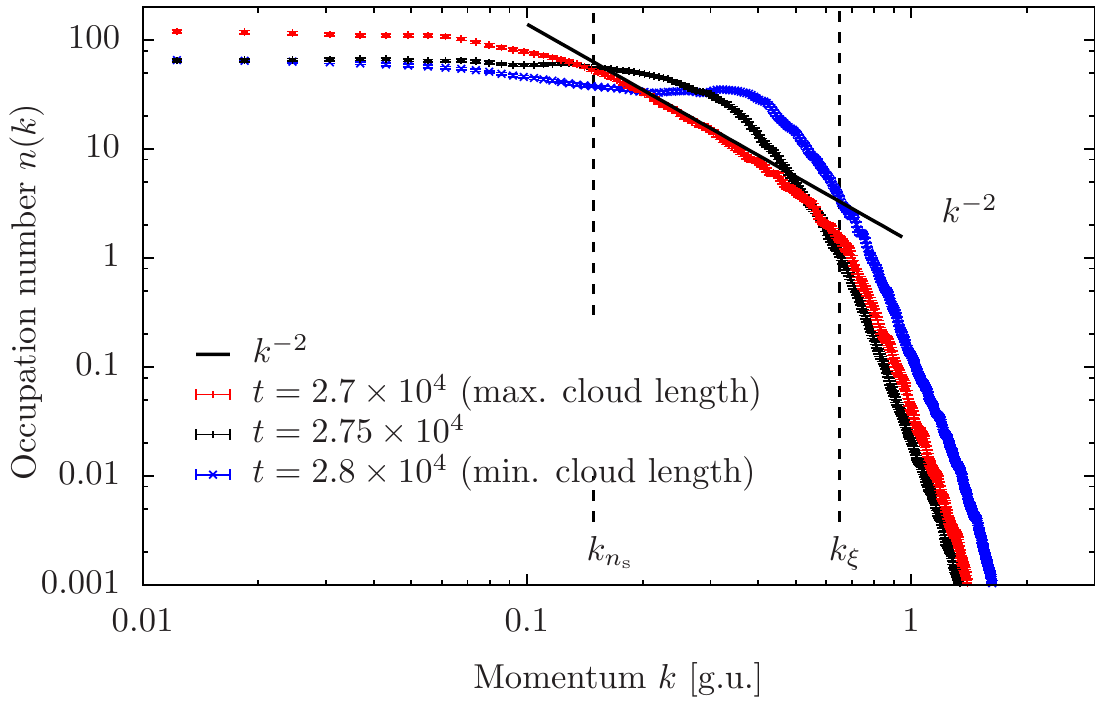}
\caption{
Momentum spectrum $n(k,t)$ at different times $t$ as indicated, with statistical errors (ensemble averages over $1000$ runs). 
Momentum scales defined by the inverse healing length, $k_{\xi}$, as well as by the density of solitons, $k_{n_{s}}$, are marked by dashed lines.
A power-law dependence $\sim k^{-2}$ is indicated by the black solid line.
Here, a larger trap oscillator length was chosen, $l_{\mathrm{ho}}=17$, such that around the outer turning point of the breathing oscillation the solitons become diluted and thus the scales $k_{n_{s}}$ and $k_{\xi}$ sufficiently far away from each other to allow for a $k^{-2}$ power law to be seen in between (red points).
Momentum $k$ and time $t$ are measured is in grid units as defined in \Sect{Simulations} where also all other simulation parameters are given. 
}
\label{fig:Spectrumk2}
\end{figure}

The single-particle momentum spectrum is determined by evaluating ensemble averages 
\begin{equation}
\label{eq:SPDMEnsemble}
n(k,t) = \langle|\phi(k,t)|^{2}\rangle_\mathrm{ensemble}
\end{equation}
over a large number of runs.
\Fig{Spectrumk2} shows $n(k,t)$ at three different times during the period where the initial breathing oscillations are still present.
Note the double logarithmic scale. 
At low momenta, the spectrum shows a plateau while at high momenta it falls of exponentially.
At the point of time corresponding to the red curve, the cloud is further expanded such that solitons are more separated from each other which causes an intermediate power law dependence $n(k)\sim k^{-2}$ to appear, as indicated by the straight line.
The plateau, the power-law and the exponential decay form the characteristic signature for solitons which we discuss in more detail in the following.
%------------------------------------------------------------------------------------------
%------------------------------------------------------------------------------------------

%==================================================================
\subsection{Random-soliton model: uniform gas}
\label{sec:RSM}
To obtain an analytical understanding of the possible spectra in the context of nonthermal fixed points we discuss, in the following, the case of a dilute ensemble of well-separated solitons with random velocities and positions.
The wave function of a single grey soliton in a homogeneous bulk background condensate is given in \Eq{SingleSoliton}. 
Using this we write down an expression for a set of $N_{s}$ uncorrelated solitons with density minima at $\{x_{i}, i=1,...,N_{s}\}$ and (dimensionless) velocities $\nu_{i}$, on a background of constant bulk density $n$:
\begin{equation}
\label{eq:NsRSfield}
\phi^{(N_{s})}(x,t)
=\sqrt{n}\prod_{i=1}^{N_{s}}\left[n^{-1/2}\phi_{\nu_{i}}(x-x_{i})\right]
\end{equation}
Note that due to the neglection of correlations this field does in general not represent a solution of the GPE in which the solitons remain non-dispersive.

We make use of the assumption that the ensemble is dilute, i.e., that the distance between each pair of neighbouring solitons is much larger than the healing length.
This assumption, for grey solitons, is not valid as soon as two oppositely moving solitons encounter each other but for simplicity we will assume that these collisional configurations can be neglected in view of a majority of well-separated solitons.
Since, for any $i$, $\phi'_{\nu_{i}}(x-x_{i})\equiv d\phi_{\nu_{i}}(x-x_{i})/dx\simeq0$ as $|x-x_{i}|\gg1$ we can rewrite the spatial derivative of the field \eq{NsRSfield} as
\begin{align}
\label{eq:dNsRSfield}
&\phi^{(N_{s})'}(x,t)
= \sum_{i=1}^{N_{s}}\phi_{\nu_{i}}'(x-x_{i},t)\prod_{j\not=i}\left[n^{-1/2}\phi_{\nu_{j}}(x-x_{j},t)\right]
\nonumber\\
&\ \ \simeq\sum_{i=1}^{N_{s}}\left[\delta(x-x_{i}(t))\prod_{j\not=i}e^{i\beta_{j}\theta(x-x_{j}(t))}\right]\star \phi_{\nu_{i}}'(x,0).
\end{align}
Here $x_{i}(t)=x_{i}(0)-\nu_{i}t$, $\beta_{j}=2\arccos\nu_{j}$, $\star$ denotes the convolution over the spatial dependence on $x$, $\theta(x)$ the Heaviside function, and we have neglected an irrelevant overall phase.
Note that the sign of $\beta_{i}$ indicates the direction of the propagation of the $i$th soliton.
The term in square brackets in the second line of \Eq{dNsRSfield} is proportional to the spatial derivative of the field describing an ensemble of $N_{s}$ infinitely thin solitons ($\xi\to0$), at the positions $\{x_{i}\}$,
\begin{align}
\label{eq:dNsRSfieldxi0}
&\phi^{(N_{s})'}(x,t)
\simeq\sum_{i}\frac{i\gamma_{i}}{2n_{s}\sqrt{n}}\phi^{(N_{s})'}_{\xi\to0}(x_{i},t)\,\delta(x-x_{i}) \star \phi_{\nu_{i}}'(x,0)
\end{align}
where $n_{s}$ is the number of solitons per unit length and the prefactor containing $\gamma_{i}$ takes into account that the phase jump by $\exp\{i\beta_{i}\theta(x-x_{i})\}$ is itself proportional to a theta function $\gamma_{i}\theta(x-x_{i})$. 
Note that although the derivative $\phi^{(N_{s})'}_{\xi\to0}(x_{i},t)$ gives a sum of terms, each being proportional to a delta distribution, only one of these remains when evaluated at $x_{i}$, which gives the term in square brackets in \Eq{dNsRSfield}.
We take the Fourier transform of $\langle{\phi^{(N_{s})'}(x)}^{*}\phi^{(N_{s})'}(y)\rangle$ with respect to $x-y$, integrate over $R=(x+y)/2$, and divide by $k^{2}$,
\begin{align}
\label{eq:FTofCoherenceFct}
&n(k,t)
= \frac{n}{4n_{s}^{2}}\Big\langle\sum_{i,j=1}^{N_{s}}{\gamma_{i}\gamma_{j}}e^{ik(x_{i}-x_{j})}\phi_{\nu_{i}}^{*}(k)\phi_{\nu_{j}}(k)
\nonumber\\
&\times\partial_{x_{i}}\partial_{x_{j}}e^{-n_{s}|x_{i}-x_{j}|\{1-\int_{\beta}P(\beta)\exp[i\beta\,\mathrm{sgn}(x_{i}-x_{j})]\}}\Big\rangle.
\end{align}
Here, $P$ is the probability for finding a soliton with greyness $\nu=\cos(\beta/2)$, and averaging over the random positions of all solitons other than those at $x_{i}$ and $x_{j}$ has been done in order to obtain the exponential decay of the coherence function.
Combining \Eq{FTofCoherenceFct} with the Fourier transform of $\phi_{\nu}(x,0)$,
\begin{align}
\phi_{\nu}(k,t)
&= i\sqrt{\frac{2\pi n}{L}}\left[{2\pi}\nu\,\delta(k)+\frac{\sqrt{2}\pi\xi}{\sinh\left({\pi\gamma k\xi}/{\sqrt{2}}\right)}\right],
\end{align}
one derives the single-particle momentum distribution for a set of $N_{s}$ solitons defined by greyness and position, $\{\nu_{i},x_{i}\,|\,i=1,\ldots,N_{s}\}$, as %(see \App{SPD})
\begin{align}
\label{eq:FullHomMomDistr}
&n(k,t)
\simeq \frac{n}{4n_{s}^{2}}\int\frac{dk'}{2\pi}\sum_{i,j=1}^{N_{s}}{\gamma_{i}\gamma_{j}}e^{i(k-k')(x_{i}-x_{j})}
\nonumber\\
&\times\left[2\pi\nu_{i}\nu_{j}\,\delta(k)+
\frac{2\delta(0)\pi^{2}\xi^{2}}{\sinh\Big({\pi\gamma_{i} k\xi}/{\sqrt{2}}\Big)\sinh\Big({\pi\gamma_{j} k\xi}/{\sqrt{2}}\Big)}
\right]
\nonumber\\
&\times\
\frac{4{k'}^{2}n_{s}\mathrm{Re}\,\alpha}{4n_{s}^{2}(\mathrm{Re}\,\alpha)^{2}+(k'-2n_{s}\mathrm{Im}\alpha)^{2}}
\,e^{-ik'(\nu_{i}-\nu_{j})t}
\nonumber\\
\end{align}
Here, the inverse volume $\delta(0)=L^{-1}$ appears as we first choose the $\phi_{\nu_{i}}^{*}$ and $\phi_{\nu_{j}}$ in \Eq{FTofCoherenceFct} different and take the identity limit only at the end.
$\alpha$ is the average over all $\alpha_{i}=(1-\exp\{i\beta_{i}\})/2$.

Assuming the dependence of $\gamma_{i}/\sinh({\pi\gamma_{j} k\xi}/{\sqrt{2}})$ on $\nu_{i}$ to be negligible we obtain an approximate stationary distribution
\begin{align}
\label{eq:ApproxStatMomDistr}
n(k)
\simeq& 
\frac{4n_{s}n \mathrm{Re}\,\alpha}{4n_{s}^{2}(\mathrm{Re}\,\alpha)^{2}+(k-2n_{s}\mathrm{Im}\alpha)^{2}}
\frac{(\pi\gamma k\xi)^{2}/2}{\sinh^{2}\Big(\frac{\pi\gamma k\xi}{\sqrt{2}}\Big)},
\end{align}
with a yet to be determined parameter $\gamma$. 
For black solitons ($\nu_{i}\equiv0$) one obtains the exact expression
\begin{align}
\label{eq:SpectrumThinBlackSolitonGas}
&\left.n(k)\right|_{\nu=0}
=\frac{4n_{s}n}{4n_{s}^{2}+k^{2}}\frac{(\pi k\xi)^{2}/2}{\sinh^{2}\Big({\pi k\xi}/{\sqrt{2}}\Big)}.
\end{align}
For an ensemble of grey solitons of identical $|\nu_{i}|\equiv\nu$, traveling with probabilities $P$ into the positive $x$-direction and $Q=1-P$ into the negative direction one finds
\begin{align}
\label{eq:SpectrumThinGreySolitonGasRandDir}
n(k)
=&\ \frac{4n_{s}n}{4n_{s}^{2}\gamma^{-4}+[k+2(1-2Q)n_{s}\nu\gamma^{-1}]^{2}}
\nonumber\\
&\times\frac{(\pi k\xi)^{2}/2}{\sinh^{2}\Big({\pi\gamma k\xi}/{\sqrt{2}}\Big)}
\end{align}
Here, $\gamma=(1-\nu^{2})^{-1/2}$.
%%
%------------------------------------------------------------------------------------------
\begin{figure}[t]
\centering \includegraphics[width=1.0 \linewidth]{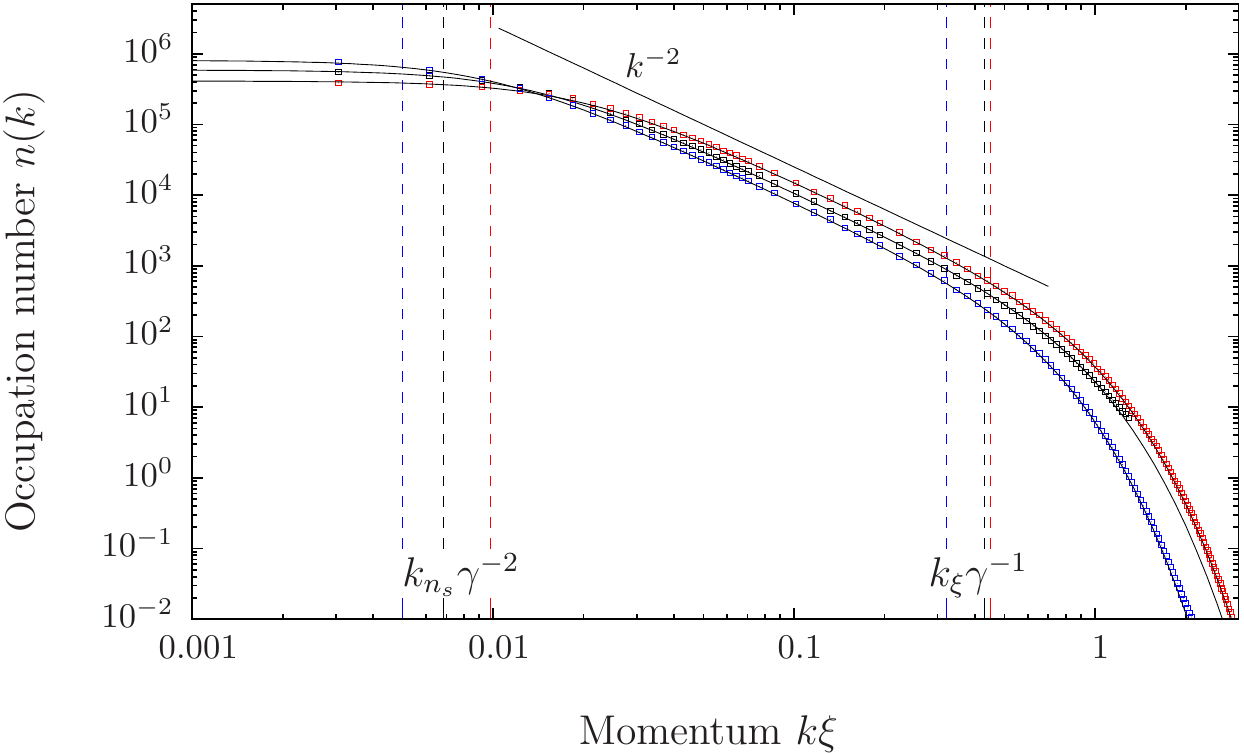}
\caption{Single-particle momentum spectrum as a function of $k$ on a double-logarithmic scale for an ensemble of $5\times10^{3}$ configurations with $N_{s}=20$ solitons each distributed according to a flat distribution across phase-space defined by the positions in the box
and the maximum greyness $|\nu_\mathrm{max}|=0.99$.
We chose $\xi=8a_{s}$.
Solid (black) squares: numerical ensemble averages, solid (black) line: \Eq{ApproxStatMomDistr} with $\alpha=0.7$, $\gamma=1.05$, shown up to $k\xi\simeq1.3$ where a finite-size effect sets in due to the random set of solitons not matching  periodic boundary conditions.
For comparison, results for $N_{s}=20$ purely black solitons (red squares and line) are shown as well as for $N_{s}=20$ solitons with fixed greyness $|\nu|=0.707$, i.e., $\gamma=1.4$, (blue squares and line) choosing an equal number of right- and left-movers, $P=Q=1/2$.
The comparison validates the approximate expressions \eq{ApproxStatMomDistr}--\eq{SpectrumThinGreySolitonGasRandDir} which exhibit scaling behaviour in the regime of momenta $k_{n_{s}}\gamma^{-2}\ll k\ll k_{\xi}\gamma^{-1}$.
}
\label{fig:ScalingSolitonsonRing}
\end{figure}
%------------------------------------------------------------------------------------------
To demonstrate the applicability of the above analytic expressions we construct ensembles of phase-space distributions of spatially well-separated solitons in a box with periodic boundary conditions and compute the ensemble average \eq{SPDMEnsemble}.
These simulations are done on a 1D grid of $N=2^{14}$ sites, generating $5\times10^{3}$ configurations for taking ensemble averages.
For this we multiply $N_{s}=20$ single-soliton solutions \eq{SingleSoliton} with positions $x_{i}$ and greyness $\nu_{i}$ chosen randomly according to a given phase-space distribution.
To make sure that their relative distance on the average is much larger than their widths we chose the phase-space distribution to allow for a maximum greyness, $|\nu_{i}|<|\nu_\mathrm{max}|<1$, such that the diluteness criterium requires an approximate minimum box length of $L=4(1-\nu_\mathrm{max}^{2})^{-1/2}N_{s}$.

\Fig{ScalingSolitonsonRing}a shows the single-particle momentum spectrum $n(k)$ on a double-logarithmic scale for an ensemble of $5\times10^{3}$ configurations with $N_{s}=20$ solitons each distributed according to a flat distribution across phase-space defined by the positions in the box 
and the maximum greyness $|\nu_\mathrm{max}|=0.99$.
Solid (black) squares represent the results of the numerical ensemble average while the solid line corresponds to the analytical formula \eq{ApproxStatMomDistr}, with fitted parameters $\alpha=0.7$, $\gamma=1.05$. 
Compare this to the analytical average $\alpha=2/3$.
For comparison, we give the results for the same number of purely black solitons (red squares and line) as well as for a fixed greyness $|\nu|=0.707$ (blue squares and line), choosing an equal number of right- and left-movers.
The comparison validates the approximate expressions \eq{ApproxStatMomDistr}--\eq{SpectrumThinGreySolitonGasRandDir} which exhibit scaling behaviour in the regime of momenta $k_{n_{s}}\gamma^{-2}\ll k\ll k_{\xi}\gamma^{-1}$.

%------------------------------------------------------------------------------------------
\begin{figure}[t]
\centering \includegraphics[width=1.0 \linewidth]{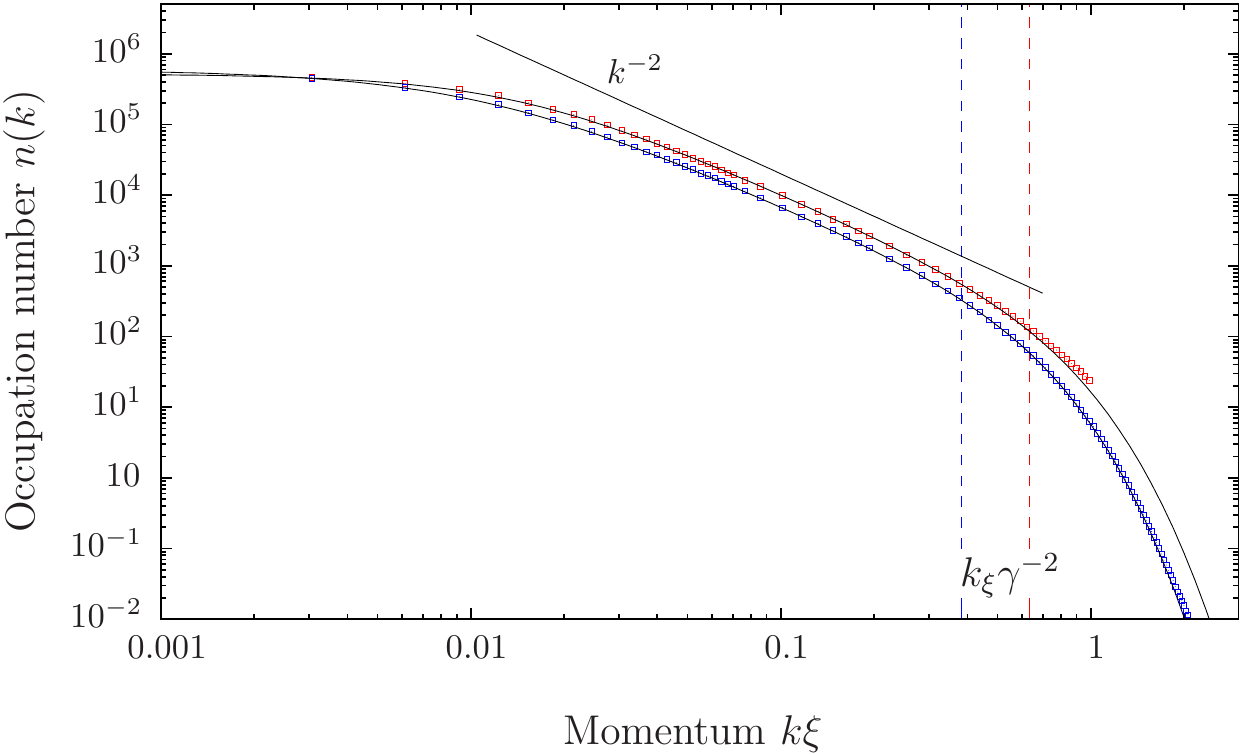}
\caption{Single-particle momentum spectrum as a function of $k$ on a double-logarithmic scale  for an ensemble of $N=5\times10^{3}$ configurations with $N_{s}=20$ solitons each distributed according to a phase-space distribution with an unequal weight for solitons with positive $\nu$ (right-movers) and negative $\nu$ (left-movers).
The greyness is uniformly distributed within $0=\nu_\mathrm{min}\le\nu\le\nu_\mathrm{max}=0.99$.
All other parameters are chosen as for \Fig{ScalingSolitonsonRing}.
Red squares and black line show the ensemble average and compare with \Eq{SpectrumThinGreySolitonGasRandDir} for $Q=0.2$, $\gamma=1.179$, and $\nu=0.412$, shown up to $k\xi\simeq1$ where a finite-size effect sets in due to the random set of solitons not matching  periodic boundary conditions.
For comparison, results for $N_{s}=20$ solitons with fixed greyness $\nu=0.707$, i.e., $\gamma=1.41$, (blue squares and line) are shown.
}
\label{fig:ScalingSolitonsonRingAsym}
\end{figure}
%------------------------------------------------------------------------------------------
In \Fig{ScalingSolitonsonRingAsym} we show the single-particle momentum spectrum for an ensemble of $5\times10^{3}$ configurations with $N_{s}=20$ solitons each distributed according to a phase-space distribution with an unequal weight for solitons with positive $\nu$ (right-movers) and negative $\nu$ (left-movers).
We specifically restricted the greyness to the interval $0=\nu_\mathrm{min}\le\nu\le\nu_\mathrm{max}=0.99$ and besides that chose the same parameters as for \Fig{ScalingSolitonsonRing}.
Black squares and line show the ensemble average and compare with \Eq{SpectrumThinGreySolitonGasRandDir} for $Q=0.2$, $\gamma=1.179$, and $\nu=0.412$.
For comparison, results for $N_{s}=20$ solitons with fixed greyness $\nu=0.707$, i.e., $\gamma=1.41$, (blue squares and line) are shown.

%==================================================================
\subsection{Relation to nonthermal fixed points and vortical excitations in superfluids}
\label{sec:NTFP}
Universal power-law behaviour in a many-body system far from equilibrium points to the appearance of turbulence phenomena.
In the following, we summarise a few basic ideas of wave-turbulence theory for quantum gases and discuss the power-law spectra derived for soliton ensembles in this context.

A dilute, degenerate Bose gas is compressible such that collective sound-wave excitations can occur.
This allows for so-called \textsl{weak} wave-turbulence which can occur in a regime where kinetic theory applies:
The wave-kinetic equation for a Bose gas, which describes the evolution of the system under the collisional interactions between different wave modes, is known to have non-trivial stationary solutions.
These solutions are nonthermal and exhibit a power-law dependence of mode occupations on momentum \cite{Zakharov1992a, Nazarenko2011a,Newell2011a}.
As in fluid turbulence, such solutions imply that energy flows from, e.g., small to large momentum scales.
In between these momentum scales of the source and the sink, the energy passes through the so-called inertial interval, where the distribution over momenta is stationary and follows a power law with a universal exponent predicted by weak-wave-turbulence theory  \cite{Zakharov1992a, Nazarenko2011a,Newell2011a}.
The stationary solution is metastable, i.e., requires a constant and equal flow in and out of the inertial interval.
We note that weak wave turbulence does not bear, in general, vortical excitations.
It is defined by scaling and stationary transport between different scales and hence can also occur in one spatial dimension where vortices are absent.

Applying the above to a degenerate Bose gas one faces, however, the problem that the description in terms of wave-kinetic equations breaks down in the infrared (IR) regime of long wavelengths where amplitudes, i.e., talking about a Bose gas, single-particle occupation numbers grow large and the description in terms of, e.g., elastic two-to-two collisions becomes unreliable \cite{Zakharov1992a}.
As a consequence of this, so-called \textsl{strong} wave turbulence is expected to occur in the IR regime.
Recent developments presented in Refs.~\cite{Berges:2008wm,Berges:2008sr,Berges:2008mr,Scheppach:2009wu,Berges:2010ez,Carrington:2010sz} allow one to set up a unifying description of scaling, both in the ultraviolet (UV) regime, where the wave-kinetic and Quantum Boltzmann equations apply,  and in the IR limit.
Recall that at a critical point the IR modes dominate the system's behaviour. 
In this IR regime, new scaling laws were found by analysing non-perturbative Kadanoff-Baym dynamic equations, as in the UV, with respect to nonthermal stationary power-law solutions ~\cite{Berges:2008wm,Berges:2008sr,Berges:2008mr,Berges:2010ez}. 
These solutions were termed nonthermal fixed points.
Analogous predictions for dilute Bose gases were given in \cite{Scheppach:2009wu}, proposing what was termed strong matter-wave turbulence in the regime of long-range excitations.

In \cite{Nowak:2010tm,Nowak:2011sk}, it was then shown that these nonthermal fixed points can also be understood, in two and three dimensions, in terms of vortex excitations of the superfluid:
In the infrared limit of large wave numbers the incompressible, superfluid component of the gas dominates, and the predicted IR power laws appear due to the algebraic radial decay of the flow-velocity around the vortex cores.
As a consequence, within a window in momentum space, which is limited by the inverse mean distance between different vortices and the inverse core size, the single-particle occupation number spectrum shows the power law predicted in \cite{Scheppach:2009wu}.

Before we discuss this in more detail let us first turn back to the soliton spectra derived in the previous section.
Assuming an equal number of solitons traveling with positive and negative velocities, $P=Q=1/2$, i.e., assuming $\mathrm{Im}\,\alpha=0$, the single-particle spectrum \eq{SpectrumThinGreySolitonGasRandDir} is characterised by a maximum of two scales.
Consider the case $n_{s}\ll\gamma/(\sqrt{2}\pi\xi)$.
For momenta greater than the reduced soliton density but smaller than the reduced inverse healing length, $k_{n_{s}}\gamma^{-2}\ll k\ll k_{\xi}\gamma^{-1}$, with $k_{n_{s}}=2n_{s}$, and $k_{\xi}=\sqrt{2}/(\pi\xi)$, the momentum distribution exhibits a power-law behaviour, $n(k)\sim k^{-2}$.
This reflects, first, the random position of the kink-like phase jump across the center of the soliton, and second, that these momenta cannot resolve the spatial width of the kink.
In other words, looking within a spatial window of size between $\gamma^{2}k_{\xi}^{-1}$ and $\gamma k_{n_{s}}^{-1}$, the appearance of a single sharp solitonic phase jump inside the window is observed in a random manner.
Hence, for any of these window sizes, the system looks identical, it appears self-similar.
This self-similarity is at the base of the scaling momentum distribution.
Already for a single soliton, the momentum distribution does not know anything about the position of the kink (this information appears as a  phase in the momentum-space Bose field which is irrelevant for $n(k)$), and thus the single-soliton distribution is self-similar, too.

For black solitons, the self-similar scaling region is limited by the scales $k_{n_{s}}$ and $k_{\xi}$. 
Below $k_{n_{s}}$, the distribution is constant because too low wave numbers cannot resolve the kink-structure.
This corresponds to the first-order coherence function decaying exponentially in space, with the decay scale set by the soliton distance $1/n_{s}$,
\begin{equation}
\label{eq:SolPhaseCohDecay}
\langle\phi^{*}({x})\phi({y})\rangle \sim \exp\{-2n_{s}|{x}-{y}|\}.
\end{equation}
Above $k_{\xi}$, the momentum spectrum resolves the finite width of the soliton density dip which results in an exponential suppression of the mode  occupations.
We recall that also in equilibrium, at sufficiently high temperatures where quasiparticle mode occupations are large, $n^{(\mathrm{qp})}(k)=k_{B}T/(c_{s}k)\gg1/2$, the first-order coherence function decays exponentially, $g^{(1)}(s)\sim\exp[-s/r_{0}]$, where the scale is set by the coherence length $r_{0}=2n/(mk_{B}T)$.
Hence, the corresponding momentum spectrum has the same shape $n^{\mathrm{eq}}(k)\sim2r_{0}^{-1}/(r_{0}^{-2}+k^{2})$ as for a set of random thin solitons.
We emphasise, however, that the transition scale $k_{n_{s}}$ above which scaling $\sim k^{-2}$ sets in (see, e.g.~\eq{SpectrumThinBlackSolitonGas}) can be made larger than for a thermal ensemble by increasing the soliton density $n_{s}$ above the inverse thermal coherence length $1/r_{0}$. 
This allows to identify non-equilibrium soliton vs.~thermal scaling in experiment.

We finally compare the universal and non-universal aspects  of the soliton momentum spectra found here with the corresponding spectra in $d=2$ and $3$ dimensions. 
As discussed in detail in Refs.~\cite{Nowak:2010tm,Nowak:2011sk}, the universal exponent $\zeta=d+2$, $d=2,3$, found for the particle spectra $n(k)\sim k^{-\zeta}$ during the dynamical relaxation of an initially strongly quenched gas reflected the appearance of vortices in two and vortex lines in three dimensions.
In two dimensions, this can be seen in an easy way looking at the flow velocity field $\mathbf{v}\sim\mathbf{\nabla}\varphi=\mathbf{e}_{\theta}/r$ at the distance $r$ from the core of a singly quantised vortex, where $\mathbf{e}_{\theta}$ is the local tangential unit vector and $\varphi$ the phase angle of the complex Bose field.
The $r$-dependence implies a $k^{-1}$ scaling of $|\mathbf{v}(k)|$ and thus a $k^{-2}$ scaling of the kinetic energy $E(k)\sim k^{2}n(k)\sim v(k)^{2}$, i.e., $n(k)\sim k^{-4}$ \cite{Nore1997b,Nowak:2011sk}.
Similar arguments lead to $n(k)\sim k^{-5}$ for the radial momentum distribution in the presence of a vortex line in three dimensions \cite{Nowak:2011sk}.
Extending these arguments, the scaling $n(k)\sim k^{-d-2}$ was shown to appear for ensembles of randomly positioned vortices/vortex lines in a range of momenta $l_\mathrm{v}^{-1}\lesssim k \lesssim\xi^{-1}$ between the inverse of the inter-vortex distance $l_{\mathrm{v}}$ and the inverse of the healing length $\xi$ which is a measure for the core width.

As pointed out above, the power-law spectrum $n(k)\sim k^{-d-2}$, in turn, had been predicted by use of nonperturbative field-theory methods in \cite{Berges:2008wm,Scheppach:2009wu} where it resulted for a strong-wave-turbulence cascade in the IR, characterising the scaling behaviour at a nonthermal fixed point.
This cascade was shown in \cite{Nowak:2011sk} to be caused by particles being transported towards the IR where they build up high mode occupations and thus coherence in the sample, see also \cite{Levich1978a,Kagan1992a,Kagan1994a,
%Kagan1997c,
Berloff2002a,Svistunov2001a,Kozik2009a}.
Note that in the picture of the evolving Bose field this momentum-space transport corresponds to the mutual annihilation of vortices and anti-vortices in the system which results in an increase of the inter-vortex distance and thus of the range over which phase coherence is established.

Having recalled all this, we note that there is a discrepancy between the predicted scaling $\sim k^{-d-2}$, which was found consistent with the vortex picture for $d=2$ and $3$, and the scaling $\sim k^{-2}$ obtained here for the solitons in $d=1$.
To gain more insight into this issue we consider the spatial decay of the phase coherence for a system in two dimensions over distances considerably larger than the intervortex spacing.
In analogy to the soliton ensembles discussed in \Sect{RSM}, we consider randomly positioned and well-separated vortices.
One finds that the decay of the phase coherence follows the same exponential law, 
\begin{equation}
\label{eq:PhaseCohDecay}
\langle\phi^{*}(\mathbf{x})\phi(\mathbf{y})\rangle \sim \exp\{-2n_{\mathrm{v,1}}|\mathbf{x}-\mathbf{y}|\},
\end{equation}
where $n_{\mathrm{v,1}}$ is the one-dimensional uniform vortex density along the straight line through $\mathbf{x}$ and $\mathbf{y}$.
Fourier transforming \eq{PhaseCohDecay} with respect to $\mathbf{x}-\mathbf{y}$ results in a momentum spectrum $n_{\mathrm{c}}(k)\sim (4n_{\mathrm{v,1}}^{2}+k^{2})^{-3/2}$ which scales as $k^{-3}$ for momenta considerably larger than the inverse vortex distance $n_{\mathrm{v,1}}=1/l_{\mathrm{v}}$.
Analogously, one finds $n_{\mathrm{c}}(k)\sim k^{-4}$ for vortex line tangles in three dimensions. 
We use the subscript `c' to distinguish these spectra from the $n(k)$ discussed above.

The apparent contradiction between the respective scalings of $n(k)$ and $n_{\mathrm{c}}(k)$ whose exponents differ by $1$, is resolved by observing the following: 
While the exponential decay \eq{PhaseCohDecay} is valid over large distances $|\mathbf{x}-\mathbf{y}|\gg l_{\mathrm{v}}$, it is the algebraic dependence of the flow field $\mathbf{v}$ as a function of distance from the vortex core which matters on length scales below the inter-vortex distance, causing the steeper power law $n(k)\sim k^{-d-2}$ at momenta $k\gg1/l_{\mathrm{v}}$.
Hence, we find an important qualitative difference between the physical properties underlying the universal scaling $n(k)\sim k^{-2}$ in one dimension and the scalings $n(k)\sim k^{-d-2}$ in $d=2$ and $3$ dimensions:
While black solitons in one dimensions are at rest and no particle flow can occur, in higher dimensions, transverse flow circling around the vortex cores gives rise to an additional contribution to the kinetic energy.
As far as this transverse flow dominates over a possible additional longitudinal flow component for which $\mathbf{v}\cdot\nabla\mathbf{v}\not=0$, see \cite{Nowak:2011sk}, this comparison also holds when allowing for grey solitons, which are moving opposite to the (longitudinal) particle flow across the soliton dip.

In summary, there is a principal difference between the scalings of the momentum spectra in one- and in higher dimensions, giving rise to a deviation of the scaling $\sim k^{-2}$ in $d=1$ from the field-theory prediction $k^{-d-2}$ which, in turn, is valid in $d=2$ and $3$ dimensions.
To recover this discrepancy within a field-theory approach to strong wave turbulence is beyond the scope of this article.

%==================================================================
\subsection{Random-soliton model: trapped gas}
\label{sec:RSM}
In our dynamical simulations we will consider soliton formation in a harmonic trapping potential rather than in a homogeneous system, see \Sect{Simulations}.
We therefore need to take into account, in the random-soliton model,  the inhomogeneous bulk distribution of the gas.

Assuming a sufficiently shallow harmonic potential, $l_\mathrm{ho}/\xi=(m\omega_\mathrm{ho}\xi^{2})^{-1/2}=(2g_\mathrm{1D}n\omega_\mathrm{ho})^{-1/2}\gg n_{s}^{-1}$ we can describe the Bose field in local-density approximation with respect to a bulk density distribution given in Thomas-Fermi approximation, $n_{\mathrm{TF}}(x)\simeq n_{0}[1-(x/R)^{2}]$, with $R=\sqrt{2}c_{s}/(\omega_\mathrm{ho}\xi)=2g_\mathrm{1D}n/\omega_\mathrm{ho}$ being the Thomas-Fermi radius in units of $\xi$.
We take the maximum density $n_{0}$ large enough to ensure $k_{\xi}\gg k_{n_{s}}$ for solitons not too close to the edge of the cloud.
In such a bulk density distribution, single solitons oscillate harmonically between classical turning points where the solitons ``touch ground'', i.e., momentarily turn black \cite{Busch2000}.
Their oscillation frequency is by a factor of $\sim1/\sqrt{2}$ smaller than the trap frequency, $\omega_{s}\simeq\omega_\mathrm{ho}/\sqrt{2}$.

In leading order in $\epsilon\sim\ddot x_{s}/\dot x_{s}$, the field of a single soliton can locally be written in the simple form given in \Eq{SingleSoliton}, with $\nu$, and thus $\gamma$ replaced by local quantities, 
\begin{equation}
\label{eq:nuinTrap}
  \nu\to\nu(x_{s})=\nu_{s,\mathrm{max}}\left[1-\left(\frac{x_{s}}{x_{s,\mathrm{max}}}\right)^{2}\right],
\end{equation}
$\gamma\to\gamma(x_{s})=[1-\nu(x_{s})^{2}]^{-1/2}$, evaluated at the position $x_{s}=x_{s}(t)$ of the soliton \cite{Busch2000}. 
$\nu_{s,\mathrm{max}}$ is the maximum greyness the soliton acquired in the center of the trap, and $x_{s,\mathrm{max}}=\nu_{s,\mathrm{max}}R$ is the distance of the soliton's turning point from the trap center.
Only solitons whose velocity does not exceed the Landau critical velocity, i.e., for which $\nu_{s,\mathrm{max}}\le1$, can oscillate in the trap for more than a quarter of the period $T_{s}=2\pi/\omega_{s}$.
This limits the maximum greyness at a distance $x$ from the trap center to a range between $0$ and $\nu_{\mathrm{max}}(x)=1-(x/R)^{2}$.

At a given time $t$, we assume a particular set $\{x_{i}(t)\}$ of $N_{s}$ well-separated solitons across the trapped gas.
The single-particle momentum spectrum corresponding to an ensemble of such sets depends on the distribution of the solitons over the greyness for each position in the trap.
This distribution is best visualised in phase space which is parametrised by the $(x,v)$, or, equivalently and in dimensionless form, by $(x/R,\nu)$, with both, $x/R$ and $\nu$ ranging between $-1$ and $1$.
In this space, the trajectory of a single soliton is a circle with radius $\nu_{s,\mathrm{max}}$ which is traced out with constant angular velocity $\omega_s$.
Hence, a stationary distribution of the $N_{s}$ solitons is given by a circularly symmetric distribution in phase space, i.e., a distribution over the different possible maximum greynesses $\nu_{s,\mathrm{max}}$ or turning points $x_{s,\mathrm{max}}$.
The simplest assumption would be that of a uniform distribution of the solitons in phase space, which amounts to a uniform distribution over the different possible $\nu$ at each distance $x$ from the trap center and an integrated soliton density distribution $n_{s}(x)\propto1-(x/R)^{2}$.

%------------------------------------------------------------------------------------------
\begin{figure}[t]
\centering \includegraphics[width=1.0 \linewidth]{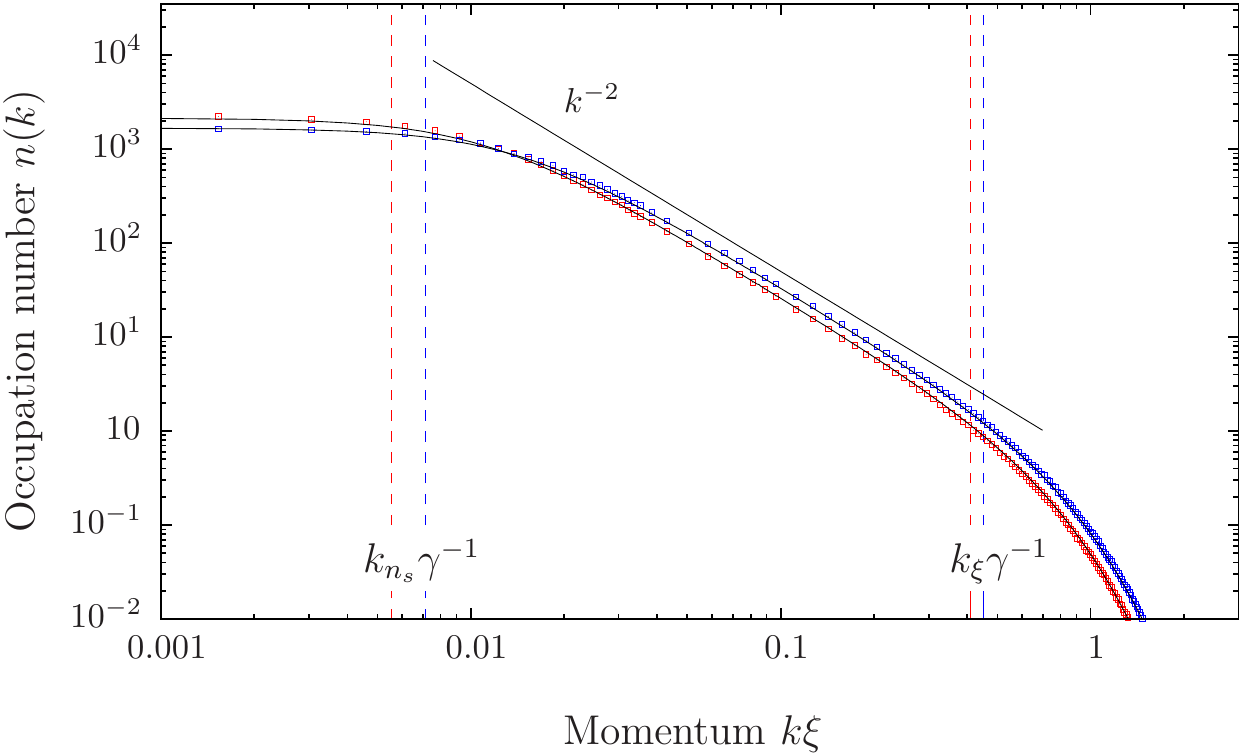}
\caption{ 
Single-particle momentum spectrum $n(k)$ on a double-logarithmic scale for an ensemble of $5\times10^{3}$ configurations with $N_{s}=20$ solitons each distributed according to a flat distribution in phase-space $\{x/R,\nu\}$, circularly symmetric around $(x=0,\nu=0)$ with radius $\bar R_{s}=1$.
We chose $\xi=4a_{s}$, $N_{0}$.
Solid (red) squares: results of numerical ensemble average; solid line through these points: analytical spectrum \eq{ApproxStatMomDistrLDA}, with $\gamma=1.1$.
(Blue) squares and line: Corresponding results for the same number of black solitons distributed randomly across the trap.
}
\label{fig:ScalingSolitonsTrap}
\end{figure}
%------------------------------------------------------------------------------------------
Following the above considerations we can obtain approximate expressions for the momentum spectrum.
For instance, for a uniform density $P[\bar x,\nu]\equiv n_{s,0}/\bar R_{s}$ of solitons within a radius $\bar R_{s}=R_{s}/R$ in phase space, i.e., for all $(\bar x,\nu)$ with $\sqrt{\bar x^{2}+\nu^{2}}\le\bar R_{s}\le 1$, with $n_{s,0}=N_{s}/(\bar R_{s}\pi)$, the first-order coherence function for thin solitons becomes
\begin{align}
&\langle{\phi^{(N_{s})}(\bar x)}^{*}\phi^{(N_{s})}(\bar y)\rangle_{\xi\to 0}
\nonumber\\
&\quad=\sqrt{n_{\mathrm{TF}}(\bar x)n_{\mathrm{TF}}(\bar y)}\exp\{-2n_{s,0}\int_{\bar y}^{\bar x}d\bar z\,\alpha(\bar z)\}
\end{align}
with a local average dephasing of
\begin{align}
\alpha(\bar x)
&=\frac{1}{2\bar R_{s}}\int_{-\sqrt{\bar R_{s}^{2}-{\bar x}^{2}}}^{\sqrt{\bar R_{s}^{2}-{\bar x}^{2}}}d\nu\left(1-\nu^{2}\pm i\nu\sqrt{1-\nu^{2}}\right)
\nonumber\\
&= \sqrt{1-\frac{{\bar x}^{2}}{\bar R_{s}}}\left(1-\frac{\bar R_{s}^{2}-{\bar x}^{2}}{3}\right).
\end{align}
Using this, the integral over $\alpha$ to linear order in $\bar x$ reads $\int_{0}^{\bar x} d\bar z\,\alpha(\bar z)=\alpha^{(1)}\bar x+\mathcal{O}(\bar x^{3})$, $\alpha^{(1)}=1-\bar R_{s}^{2}/3$. 
This approximation is best for $\bar R_{s}\to 1$, in which limit we obtain
\begin{align}
&\langle{\phi^{(N_{s})}(\bar x)}^{*}\phi^{(N_{s})}(\bar y)\rangle_{\xi\to 0}
\nonumber\\
&\quad
\simeq\sqrt{n_{\mathrm{TF}}(\bar x)n_{\mathrm{TF}}(\bar y)}e^{- 4N_{s}|{\bar x}-{\bar y}|/(3\pi\bar R_{s})}.
\end{align}
Analogously, one calculates the exponential dephasing factor for more complicated soliton phase-space distributions.
Taking the above results together one derives the momentum distribution in local-density-approximation as the convolution of the spectrum for a homogeneous distribution of thin solitons with the Fourier transform of the bulk density, multiplied with the momentum spectrum of a single soliton:
\begin{align}
\label{eq:ApproxStatMomDistrLDA}
n(k)
\simeq& \left(n_{\mathrm{TF}}(k)\star
\frac{4n_{s,0}\alpha^{(1)}}{4n_{s,0}^{2}\alpha^{(1)2}+k^{2}}\right)
\frac{(\pi\gamma k\xi)^{2}/2}{\sinh^{2}\Big({\pi\gamma k\xi}/{\sqrt{2}}\Big)},
\end{align}
where the parameter $\gamma$ is to be determined and $\star$ denotes the convolution with respect to $k$.

In the case that single soliton distributions contributing to the ensemble are not uniform throughout phase space, the ensemble-averaged $n(k)$ would rather be a sum of ($k\to-k$)-asymmetric distributions such that on the average the momentum distribution can have local maxima at finite $|k|>0$.

To study the quality of the above analytic expressions we construct ensembles of phase-space distributions of spatially well-separated solitons inside a harmonic trap and compute the ensemble average \eq{SPDMEnsemble}.
For this we multiply $N_{s}$ single-soliton solutions \eq{SingleSoliton} with positions $x_{i}$ and greyness $\nu_{i}$ chosen according to a given phase-space probability distribution and ensure that their relative distance on the average is much larger than their widths.

\Fig{ScalingSolitonsTrap} shows the single-particle momentum spectrum $n(k)$ on a double-logarithmic scale for an ensemble of $5\times10^{3}$ configurations with $N_{s}=20$ solitons each distributed according to a flat distribution in phase-space $\{x/R,\nu\}$, circularly symmetric around $(x=0,\nu=0)$ with radius $\bar R_{s}=1$.
Solid (red) squares represent the results of the numerical ensemble average while the solid line corresponds to the analytical formula \eq{ApproxStatMomDistrLDA}, with $\gamma=1.1$.
For comparison, we give corresponding results for the same number of black solitons distributed randomly across the trap (blue squares and line).

%==================================================================
\section{Soliton spectra in dynamical simulations}
\label{sec:Simulations}
%
%==================================================================
\subsection{Soliton formation and tracking in position space}
\label{sec:realspace}
In the following we study the formation of soliton ensembles by use of semiclassical simulations, with Gaussian noise for the initial field modes.
Moments of the phase-space probability distribution at a later time are determined by sampling the initial distribution, propagating each realisation according to the GPE, and averaging over many such trajectories  \cite{polkovnikov2010,blakie2008}.
At the initial time, we take the gas to be noninteracting and thermalised and impose an interaction quench.
To allow the emerging collective excitations to form solitons at a desired density we furthermore apply evaporative cooling by opening the trapping potential at the edges in a controlled fashion.
During the first, cooling period, $t \le t_{\mathrm{c}}$, the potential is given by the inverted Gaussian $V(x,t) = m\omega_\mathrm{ho}^{2}U(t)\{1-\exp[-x^2/2U(t)]\}$ with its maximum being ramped down by sweeping $U(t)=U_{0}+(U_{c}-U_{0})t/t_{c}$ linearly in time from $U_{0}$ to $U_{c}$.
At the same time, highly energetic particles near the edge of the potential are removed by adding a loss term $\i \Gamma(x,t)/2= i\Gamma_{\infty} [V(x,t)/U(t)]^{r}/2$ to the trapping potential. 
Thereafter, during the interval $t_{\mathrm{c}} \le t \le t_{\mathrm{max}} $ the loss is switched off and the potential is ramped up again to harmonic shape accross the extension of the gas, $U(t)=U_{c}+(U_\mathrm{max}-U_{c})(t-t_{c})/(t_{\mathrm{max}}-t_{c})$.
We choose $r=10$, $U_{0}=2.75$, $\Gamma_{\infty}=0.1\,U_{0}$, $U_{c}=U_{0}/3$, and $U_{\mathrm{max}}=10\,U_{0}$.
The times $t_{c}$ and $t_{max}$ vary and are given in the following.
This protocol corresponds to the one used in Ref.~\cite{Witkowska2010}.
Different cooling schemes have been used in experiments, see, e.g., \cite{hofferberth2007, Kitagawa2011}, but as we are primarily interested in the one-dimensional dynamics, we here restrict ourselves to purely 1D calculations.

For the simulations, we map the system onto a grid of $N=1024$ lattice sites with a lattice constant $a_{s}$.
If not stated otherwise, quantities are given in grid units based on $a_{s}$, and the parameters chosen are a dimensionless coupling constant $\overline g_{\mathrm{1D}}= 2 m a_{s} g_{\mathrm{1D}} = 7.3\times10^{-3}$, a cooling time $\overline t_{\mathrm{c}} =t_{c}/(2ma^2) = 9.1\times10^{3}$, and a harmonic oscillator length $\overline l_{\mathrm{ho}} =   l_{\mathrm{ho}}/a = 8.5$.
Lattice momenta are $\overline k_{n}= 2 \sin(\pi n/{N)}$, $n\in \{-N/2,...,N/2-1\}$.
We drop overbars in the following.

Three stages of the induced dynamical evolution can be  observed, see also \Fig{RSPACEevolution}:
%------------------------------------------------------------------------------------------
\begin{figure}[t,b,p]
\centering \includegraphics[width=0.45\textwidth]{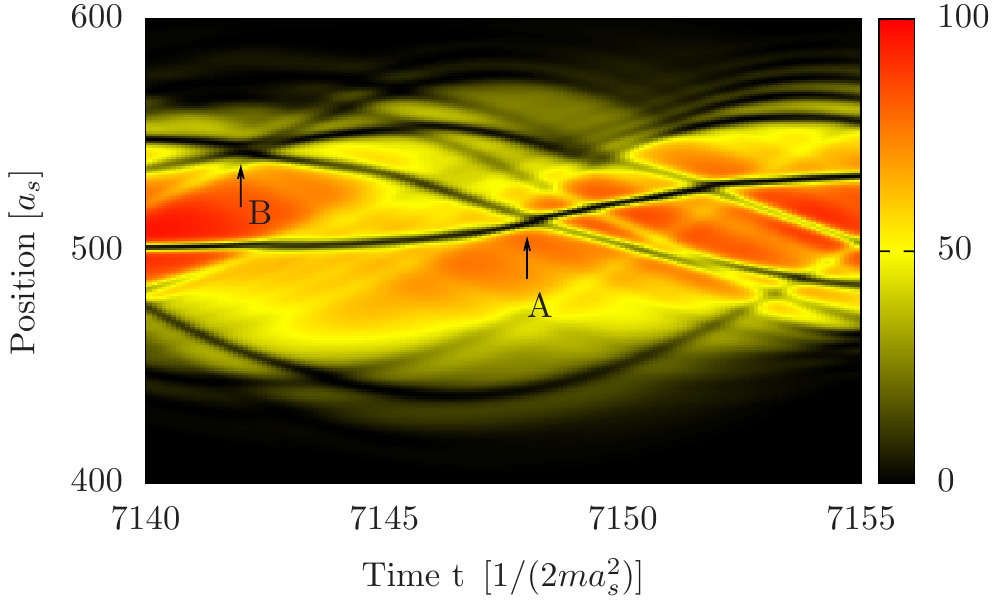}
\caption{Snapshots of a single run of the non-linear classical field equation, showing solitons which oscillate inside the trap, thereby showing signs of mutual scattering and passing through each other (examples marked by letters).
Parameters are chosen as in \Fig{RSPACEevolution}.}
\label{fig:SolitonExample}
\end{figure}
%------------------------------------------------------------------------------------------

\textit{Initial Oscillations.}
See the top left panel of \Fig{RSPACEevolution}. 
Following the interaction quench potential energy is transferred to kinetic energy. 
One observes strong breathing-like oscillations of the gas. 
These oscillations decay on a timescale of $t \approx 2\times10^{3}$, leaving a dipolar oscillation of the bulk in the harmonic trap. 
Solitons are formed in the wake of the decaying breathing oscillations.

\textit{Solitonic Regime.}
See the top right panel of \Fig{RSPACEevolution}. 
The initial collective oscillations have largely decayed except for an overall dipole mode, and many solitons appear.
The bottom left graph shows the evolution around $t=t_{\mathrm{c}}=9.1\times10^{3}$ when only very few solitons have survived. 
The solitons oscillate in the trap, being nearly black at the edges and grey in the center of the trap corresponding to a nonzero velocity.
\Fig{SolitonExample} magnifies a short period of the evolution.
On mutual encounters, the solitons get phase-shifted, such that collisions show signs of scattering or passing through each other.
Collisions with different such shifts are marked by letters A and B in \Fig{SolitonExample}.

\textit{Final stage:}
At times $t \gg t_{\mathrm{c}}$, a soliton is still visible, see the bottom right panel of \Fig{RSPACEevolution}. 
Comparing runs we find different numbers of solitons remaining during the late stage.

The smallest time scale is the oscillation period in the trap $T_{\mathrm{ho}} \approx 230$, which leads to an initial collective oscillation with period $T_{\mathrm{oscillation}}  \approx 300$ (cf.~\Fig{RSPACEevolution}). 
The collective breathing motion dies out after $\tau  \approx 2000$. 
The oscillation period of a soliton in the Thomas Fermi bulk is $T_s \approx \sqrt{2} \, T_{\mathrm{ho}} \approx 320$. 
The longest time scale in our setup is the cooling time $t_{\mathrm{c}} = 9100$. 
Comparing these time scales to the total time of the simulation, at the end of which solitons are still present, we see that the solitons are quasi-stationary in the system. 
They emerge soon after the initial quench and remain throughout the whole evolution while thermalisation of the high-momentum modes is proceeding as we will see in the following.

%------------------------------------------------------------------------------------------
\begin{figure}[t]
\centering \includegraphics[width = 1.0 \linewidth]{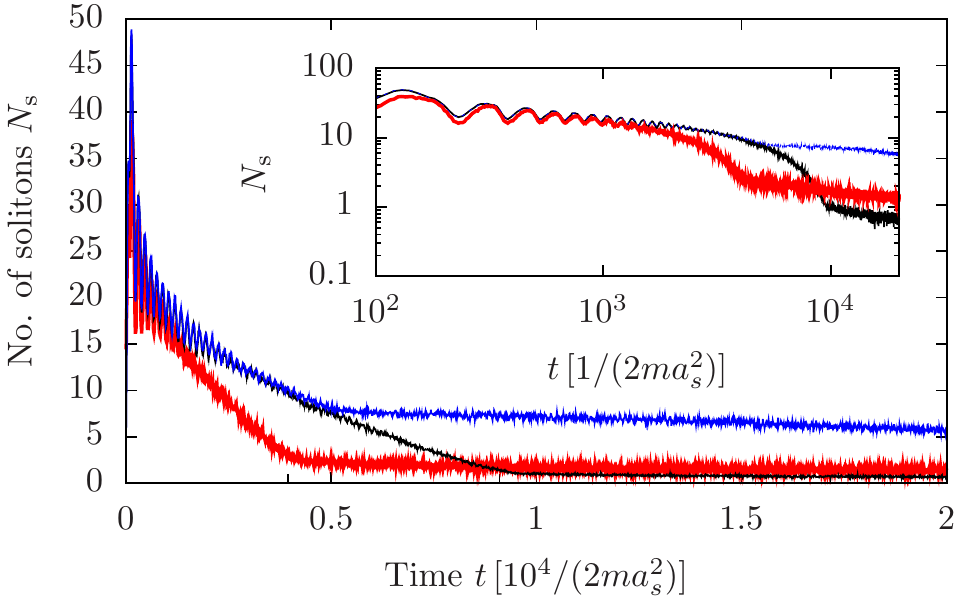}
\caption{Time evolution of the mean number $N_{s}$ of solitons, for an ensemble of $100$ runs. 
Strong initial oscillations occur during the breathing like bulk oscillations after the interaction quench, until $t\simeq 10^{3}$. 
The number of solitons decreases while the gas is evaporatively cooled.
After the end of the cooling, at $t=t_{c}$  the decay is considerably slowed down and the number of solitons remains largely stable.
Three different cooling times and two ramp speeds are shown, with $t_{c}=3.9\times10^{3}$ (red), $t_{c}=4.55\times10^{3}$ (blue), and $t_{c}=9.1\times10^{3}$ (black).
}
\label{fig:SolNumber}
\end{figure}
%------------------------------------------------------------------------------------------
%==================================================================
\subsection{Number of solitons after cooling ends}
In order to study the statistics of the solitons emerging during the evolution we have set up an efficient tracking algorithm which identifies the trajectories of the solitons oscillating in the gas.
The algorithm scans the wave function for density minima coinciding with a phase jump around them.
\Fig{SolNumber} shows the evolution of the mean number of solitons, for an ensemble of $200$ runs. 
The three stages described above can be identified. 
The strong initial oscillations give an oscillating number of solitons until $t\simeq 10^{3}$. 
The number of solitons decreases while the gas is evaporatively cooled.
After the end of the cooling, at $t=t_{c}$  the decay is considerably slowed down and the number of solitons remains largely stable.
Three different cooling times and two ramp speeds are shown, with $t_{c}=3.9\times10^{3}$ (red), $t_{c}=4.55\times10^{3}$ (blue), and $t_{c}=9.1\times10^{3}$ (black), where the same speed is chosen to obtain the blue and black data.\\

\begin{figure}[t,b,p]
\centering \includegraphics[width = 0.8 \linewidth]{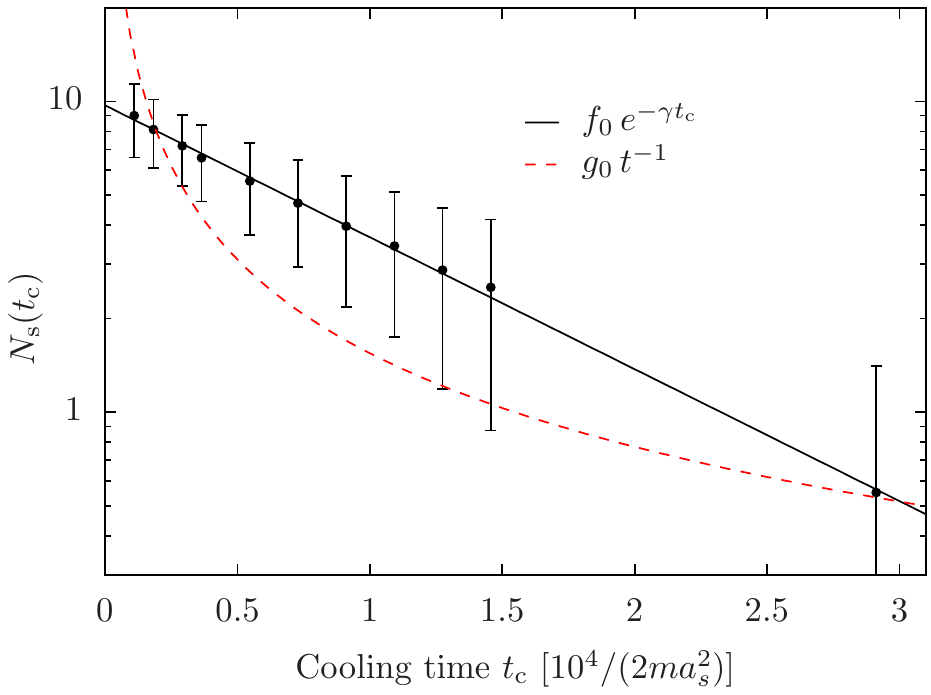}
\caption{The number of solitons $N_{s}(t_{c})$ at the end of the cooling period, $t=t_{c}$,  as a function of $t_{c}$, for ensembles of 200 runs each (black dots and error bars). 
The function $f_{0}\exp\{-\gamma t_{c}\}$, with $f_0=9.7$ and $\gamma=9.77 \cdot 10^{-5}$, was fitted with $\chi^2 = 0.006$ (black line). 
The function $g_{0}t_{c}^{-1}$, with $g_0 = 1.55 \cdot 10^5$, was fitted with $\chi^2 = 1.28$ (red dashed line).
Hence, other than for the Kibble-Zurek scheme of Ref.~\cite{Witkowska2010}, cooling after the initial quench results in an exponential dependence of $N_{s}(t_{c})$ on $t_{c}$.
}
\label{fig:CompareNumSol}
\end{figure}

Kibble and Zurek have predicted that the number of defects created in the near-adiabatic crossing of a phase transition scales with the crossing rate according to a power law which depends on the universal properties of the transition \cite{kibble1976, zurek1985}.
This was studied numerically in \cite{Witkowska2010} using the cooling protocol described above.
While the interacting gas was chosen to be in thermal equilibrium initially, with a temperature well above the critical point, we start our simulations, motivated by earlier work on vortex dynamics \cite{Nowak:2010tm,Nowak:2011sk}, with an interaction quench driving the system strongly out of equilibrium.
To compare the dynamics induced in this way with the results of \cite{Witkowska2010} we show, in \Fig{CompareNumSol}, the dependence of the number of solitons created on the cooling ramp time $t_{c}$.
We find that, within the error bars which indicate the variance over $200$ runs, the data is rather fitted by an exponential dependence $N_{s}(t_{\mathrm{c}})=f_{0}\exp\{-\gamma t_{\mathrm{c}}\}$ than by a power law $N_{s}(t_{\mathrm{c}})= g_{0}/t_{\mathrm{c}}$ as predicted in Ref.~\cite{Zurek2010}. 
We emphasise however that in our system, solitons mainly form during the initial stage following the interaction quench.
%------------------------------------------------------------------------------------------
%------------------------------------------------------------------------------------------

%==================================================================
\subsection{Time evolution of single-particle spectra}
\begin{figure}[tb]
\centering \includegraphics[width=0.9 \linewidth]{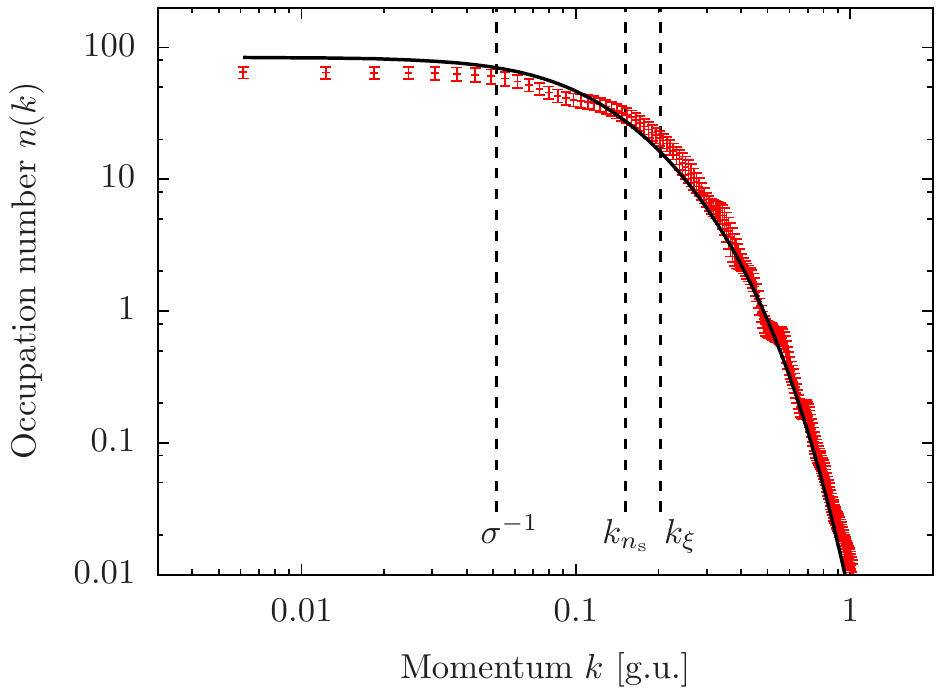}
\caption{
Momentum spectrum $n(k,t)$ at time $t=5\cdot10^{3}$ with statistical errors (ensemble average over $100$ runs).
$k$ and $t$ are in grid units as defined in \Sect{realspace} where also all other simulation parameters are given.  
Solitons have formed at high density such that the scales $k_{n_{s}}$ and $k_{\xi}$ are close together, as indicated in the graph.
Solid line: fit of the analytical model spectrum \eq{ApproxStatMomDistrLDA}, with $n_{s,0} = 0.076$,  $\gamma=1$, $\sigma^-1 = 0.036$ where Gaussian of width $\sigma$ was used to describe the bulk distribution in position space.
A Rayleigh-Jeans tail is absent as the cooling is still on.
Due to the proximity of $k_{n_{s}}$ and $k_{\xi}$ no $k^{-2}$ power law is seen in between the low-energy plateau and the high-energy exponential fall-off.
}
\label{fig:SpectralFit}
\end{figure}

We finally discuss the relaxation dynamics with respect to the evolution of the respective single-particle momentum spectra \eq{SPDMEnsemble}.
The initial state chosen in the simulations is given by a thermal canonical ensemble of distributions over the single-particle eigenstates of the trap.
In \Fig{SpectralFit} we show the momentum spectrum at time $t=5\cdot10^{3}$.
Solitons have formed at high density such that the scales $k_{n_{s}}\simeq0.15$ and $k_{\xi}\simeq0.2$ are close together, as indicated in the graph.
The solid line represents a fit of the analytical model spectrum \eq{SpectrumThinGreySolitonGasRandDir}.
A Rayleigh-Jeans tail is absent as the cooling is still on. 
Due to the proximity of $k_{n_{s}}$ and $k_{\xi}$ no $k^{-2}$ power law is seen in between the low-energy plateau and the high-energy exponential fall-off.
%------------------------------------------------------------------------------------------
%------------------------------------------------------------------------------------------
%------------------------------------------------------------------------------------------
\begin{figure}[tb]
\centering \includegraphics[width = 0.9 \linewidth]{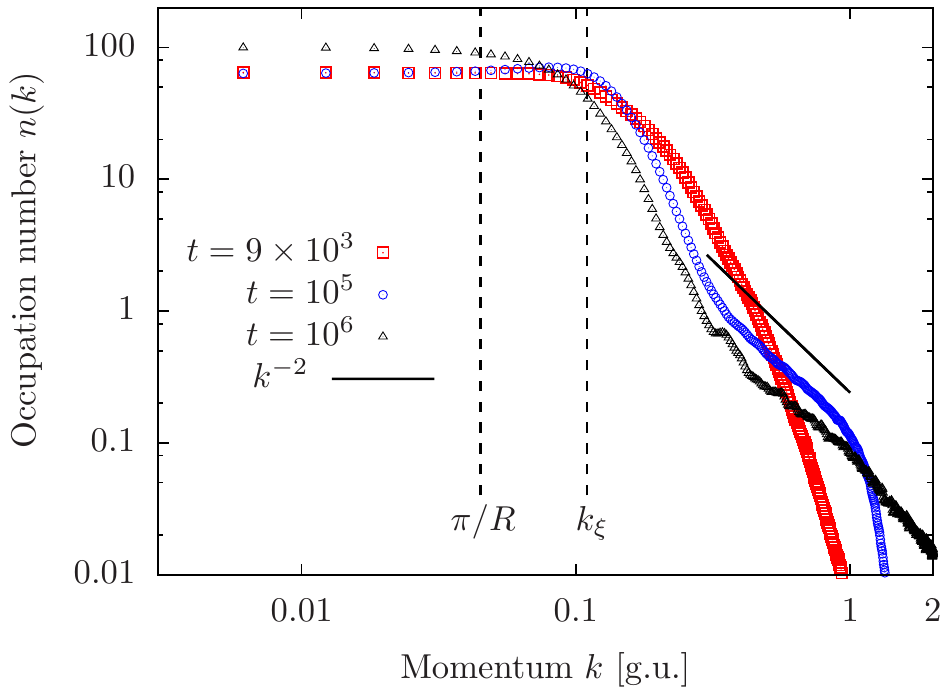}
\caption{The spectrum $n(k)$ is shown for three different times from $t \approx t_{\mathrm{c}}=9\times10^{3}$ to $t =10^{6}$, for an ensemble of $5\times10^{3}$ ($10^{3}$ for $t=10^{6}$) realisations. 
$k$ and $t$ are in grid units as defined in \Sect{realspace} where also all other simulation parameters are given. 
Momentum scales defined by the inverse Thomas-Fermi radius $R$ as well as by the healing length $\xi$ are marked by dashed lines.
A power-law dependence $\sim k^{-2}$ of the thermal tail is indicated by the black solid line.
}
\label{fig:FinalSpectra}
\end{figure}
%------------------------------------------------------------------------------------------

%------------------------------------------------------------------------------------------
\begin{figure}[tb]
\ \\
\centering \includegraphics[width =  \linewidth]{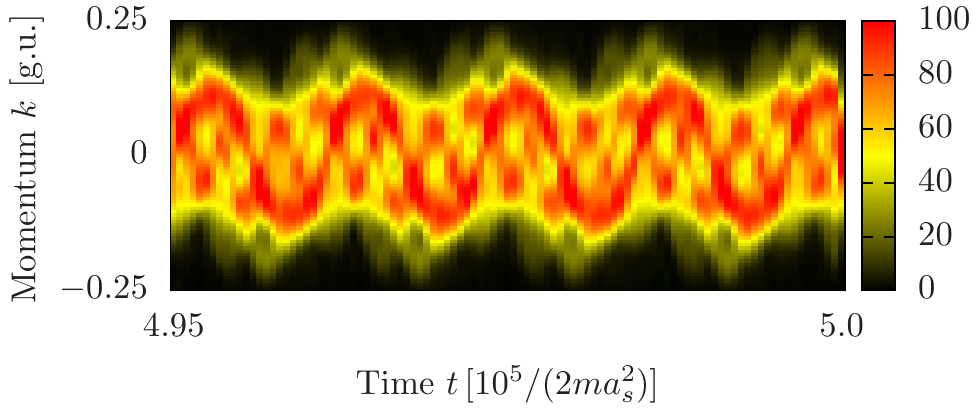}
\caption{
Time evolution of the spectrum shown on a linear color scale at late time $4.95<10^{-5}t<5.0$, for a single run.
The oscillatory pattern appears despite a large number of runs contributing to the statistical ensemble.
A small number of runs with a few oscillating solitons dominate the averaged spectrum. 
}
\label{fig:KSPACEMaxima}
\end{figure}
%------------------------------------------------------------------------------------------
In \Sect{SPMS}, \Fig{Spectrumk2} we showed the spectrum for a wider trap with $l_{\mathrm{ho}}=17$ which allows the solitons to be diluted more across the trap and results in $k_{n_{s}}\simeq0.15$ and $k_{\xi}\simeq0.65$.
This allows for a $k^{-2}$ scaling to appear clearly, indicating a self-similar random distribution of the solitons.
For $t>t_{\mathrm{c}}$, the gas enters the final stage: 
The number of particles and the energy are now conserved and the healing length is `frozen out'. 
\Fig{FinalSpectra} shows the development of $n(k,t)$ from $t \gtrsim t_{\mathrm{c}}$ to late times. 
Once the cooling and thus the removal of particles with high energy is terminated, a transport process from low to high momenta starts and thermalisation takes place. 
The influence of solitons is still present with two effects: 
First, there are still solitons in the gas for the times displayed in \Fig{FinalSpectra} which contribute with their spectral profile to the total spectrum and broaden the plateau at low momenta up to $k_{\xi}\simeq0.1$. 
Second, the momentum distribution starts to oscillate between situations with a stronger weight on the positive and on the negative side.
\Fig{KSPACEMaxima} shows how the remaining solitons, which oscillate in the trap, influence the spectrum with their own momentum appearing as the oscillating maxima in the spectrum.

%==================================================================
\section{Conclusions}
\label{sec:Conclusions}
We have studied the formation of dark solitary waves in one-dimensional Bose-Einstein condensates as well as their relaxation dynamics towards equilibrium.
The corresponding single-particle momentum spectra were predicted in the framework of an instantaneous model of well-separated grey solitons, whose width is considerably smaller than their mutual distances in the bulk.
For comparison with these predictions, semiclassical simulations of the relaxation dynamics of one-dimensional Bose gases after an initial interaction quench and a cooling period were used to determine the respective spectra numerically.
The so found spectra compared well with the analytical predictions, giving insight into the many-body dynamics from the point of view of universal properties and critical physics far from equilibrium.
We emphasise that the particular protocol used to produce the solitons is irrelevant in that the properties characterising the fixed point do not depend on how it is reached.

We have discussed the power-law behaviour of the momentum spectra, which appears in a range of momenta between the inverse of the inter-soliton distance and the inverse healing length, with regard to the universal scaling laws predicted in non-perturbative field-theory approaches to strong wave turbulence.
In the one-dimensional case studied here, the derived power-law exponent $\zeta=2$ differs by 1 from the exponent $\zeta=d+2$ predicted for a system in $d$ spatial dimensions and previously recovered in the scaling due to  vortex excitations in a $d=2$ and $3$-dimensional Bose gas.
We trace this discrepancy back to the different flow patterns possible in $d>1$ versus $d=1$ dimensions:
While in the one-dimensional gas, particle flow cannot choose its orientation except for a sign, vortical excitations in two and three dimensions are characterised by flow circling around the vortex cores.
This flow is transverse, i.e., it changes its strength perpendicular to its direction. 
It dominates, at sufficiently low energies and momenta, over any additional longitudinal flow caused by compressible sound-wave excitations.
Moreover, for geometric reasons, the transverse-flow velocity field changes algebraically in space and causes the single-particle momentum spectra to scale as predicted in strong-wave-turbulence theory.

We point out that the single-particle momentum spectra discussed here could be used in experiment to study solitary-wave dynamics in one-dimensional Bose gases without the necessity to detect solitons in situ.
Studying in this way universal properties during the relaxation dynamics from a non-equilibrium initial state or under a constant driving force opens a new access to strong wave turbulence and nonthermal fixed points.

%==============================================================================

\acknowledgments
The authors thank J.~Berges, R.~B\"ucker, L.~Carr, M.~J.~Davis, M.~Karl, G.~Nikoghosyan, M.~K.~Oberthaler, J.~M. Pawlowski, J.~Schmiedmayer, and J.~Schole for useful discussions. 
They acknowledge support by the Deutsche Forschungsgemeinschaft (GA 677/7,8), by the University of Heidelberg (FRONTIER, Excellence Initiative, Center for Quantum Dynamics), and by the Helmholtz Association (HA216/EMMI).

%==================================================================
%==================================================================
%\bibliography{MaxBib,bibtex/mybib,bibtex/additions}

\end{document}